\documentclass[aps,twocolumn]{revtex4}
\usepackage{epsfig}
\usepackage{feynmp}
\newcommand{\be}{\begin{equation}}
\newcommand{\ee}{\end{equation}}
\newcommand{\ba}{\begin{eqnarray}}
\newcommand{\ea}{\end{eqnarray}}

\def\GeV{\ {\rm GeV}}
\def\GeV2{{\rm\ GeV}^2}

\begin{document}

\title{Electromagnetic structure functions and neutrino nucleon scattering}
\author{M. H. Reno}
\affiliation{%
Department of Physics and Astronomy, University of Iowa,
Iowa City, Iowa 52242 USA}
\begin{abstract}

Electromagnetic structure functions for electron-proton scattering
are used as a test of the QCD improved parton model 
at low and moderate $Q^2$. 
Two parameterizations which work well in $ep$ scattering at low
$Q^2$ 
are used to evaluate the inelastic muon neutrino-nucleon 
and muon antineutrino-nucleon cross sections for
energies between 1-10 GeV, of interest in long baseline
neutrino oscillation experiments. Cross sections are reduced
when these low-$Q^2$ extrapolations are used.

\end{abstract}

\maketitle

\section{Introduction}

Neutrino physics has been a topic of considerable interest, especially
since the discovery of neutrino oscillations \cite{maltoni}
in the context of
solar neutrinos \cite{solar} and atmospheric neutrinos
\cite{atm}. Determinations of neutrino
mass squared differences and mixing angles have led to long baseline
neutrino experiments \cite{accel} 
where the energies of the neutrino beams
range between a few hundred MeV to tens of GeV. In the few GeV region,
it is a theoretical challenge to 
describe the neutrino-nucleon cross section
with high precision \cite{lipari}. 
Theoretical issues include the problem of making the transition
between exclusive and inclusive calculations, and the fact that one
is generally in a $Q^2$ regime where the coupling constant
$\alpha_s(Q^2)$ is large. 

Currently, experimental data in this kinematic region for
neutrino scattering are sparse \cite{zeller}, although experiments are
planned to remedy this situation \cite{minerva,finesse}.
Theoretical work has started to address the issue of combining
exclusive and inclusive processes in Ref. \cite{new}. One
element of the calculation is the deep inelastic scattering (DIS)
contribution, which is the focus of this paper.
We assess the perturbative QCD description of the electroweak structure functions in the kinematic regime relevant to
the neutrino cross section at a few GeV incident energy.
We do this by considering 
the electron-proton electromagnetic structure functions
where there are extensive data \cite{pdg}. The breakdown
of the perturbative description occurs at a $Q^2$ scale of
$Q^2\sim 1$ GeV$^2$ when one compares electromagnetic scattering
data with structure function 
calculations. In electromagnetic scattering,
below $\sim 1 \GeV2$, phenomenological parameterizations can be
used in place of the parton model based evaluations of the structure
functions. 

There are parameterizations of the structure
functions over the full
range of $x$ and $Q$ that successfully describe the electromagnetic
data \cite{allm,others}, 
however, it is not completely obvious how to generalize these
parameterizations to the neutrino scattering case. 
Bodek, Yang and Park \cite{by,byp} have taken a different approach,
namely to extract flavor components of structure
functions even in the nonperturbative regime \cite{petti}. 
This is explicitly
applicable to neutrino scattering.
Here, we examine a structure function
parameterization 
by Capella et al.  \cite{ckmt,ckmtnu}
which does well in electron-proton scattering
at low values of $Q^2$
and at the same time has a straightforward transformation to
neutrino charged current scattering. The parameterizations of
Capella et al., 
and by Bodek, Yang and Park 
are described and compared below. 

In the next section, we show the results of a perturbative evaluation
of the DIS charged current cross section in the intermediate
(few GeV to 10 GeV) energy range to establish which kinematic
regions contribute most to the cross section. In Section III,
we use $ep$ scattering results to show the range of
validity of perturbative QCD and the efficacy of the two phenomenological
parameterizations 
of the structure functions into the non-perturbative
regime. Section IV shows how the extrapolations translate to the
structure functions relevant to
neutrino scattering. Comparisons of the
different approaches to
low $Q^2$ structure functions in neutrino-nucleon charged current
cross sections are shown in Sec. V.

\section{DIS in $\nu N$ scattering}

Of particular interest is the neutrino cross section
for energies up to 10-20 GeV. One approach \cite{lipari}
to the cross section in this
energy regime is to add three separate contributions to the
cross section: the (quasi)elastic weak scattering contribution which
leaves the nucleon intact \cite{qe}, the
resonant contribution in which a finite number of resonances including the $\Delta$ 
are included \cite{delta}, and the inelastic contribution to sum the
remaining terms \cite{dis}. 
To avoid double counting, the evaluation of
the inelastic piece is done over a restricted phase space.  Generally,
something like a limit on the hadronic final state invariant
mass  $W$, such $W^2>W^2_{\rm min}\sim 2$ GeV$^2$, is applied.
We use $W_{\rm min}$ to separate the exclusive and inclusive
calculations, and we focus only on the inclusive portion of the
cross section.

As Lipari, Lusignoli and Sartogo emphasized \cite{lipari}, 
the charged current cross section components 
for $\nu_\mu N$ scattering from quasi-elastic, $\Delta$ resonance 
production, and deep inelastic scattering (DIS) with $W^2>2$ GeV$^2$ are about equal
for $E_\nu\sim 2$ GeV. The DIS term grows with increasing energy.

\begin{figure}[t]
\begin{center}
\epsfig{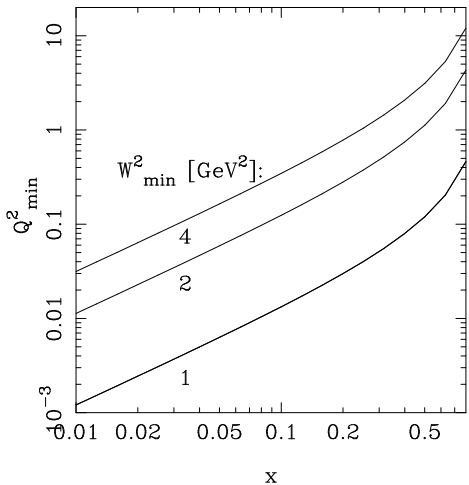}
\end{center}
\caption{Minimum value of $Q^2$ such that $W^2>W_{\rm min}^2=1,\
2$ and 4 GeV$^2$. 
}
\label{fig:wcontours}
\end{figure}

``Deep'' inelastic scattering is a misnomer in this case
because of the sensitivity to the cross section to low-$Q^2$.
Neutrino-isoscalar nucleon scattering 
$$
\nu_\mu (k)\, N(p)\rightarrow \mu(k')\, X
$$
is discussed in terms of $q=k-k'$, $Q^2\equiv -q^2\geq
0$, the invariant momentum transfer to the hadronic
system, and $x\equiv Q^2/(2p\cdot q)$. The nucleon mass is labeled
with $M$.
Kinematics show that the hadronic final state invariant mass is
\begin{equation}
 W^2=Q^2\Biggl( \frac{1}{x}-1\Biggr) +M^2\ ,
 \label{eq:w2}
\end{equation}
so $Q$ is kinematically allowed to range
in the non-perturbative regime. Fig. 
\ref{fig:wcontours} shows the minimum values of $Q^2$ such
that $W^2>W_{\rm min}^2=1,\
2$ and 4 GeV$^2$.
The contribution of the low-$Q^2$ kinematic region to the 
DIS neutrino cross section is the focus of this paper.

The neutrino differential cross section can be written in
terms of structure functions $F_i$ as
\begin{eqnarray}
\frac{d^2\sigma_{CC}^\nu(\nu N)}{dx\, dy} & = & 
\frac{G_F^2 M E_\nu}{\pi (1+Q^2/M_W^2)^2}
\Biggl[xy^2 F_1^{TMC}\nonumber \\
&+& \Biggl( 1-y-\frac{Mxy}{2E_\nu}\Biggr) F_2^{TMC}\nonumber \\
&+&\Biggl(xy-\frac{xy^2}{2}\Biggr) F_3^{TMC}\Biggr]
\end{eqnarray}
when the outgoing lepton mass is neglected. The full expression
including lepton masses is found in Ref. \cite{kr}, for example.
Here, we assume $\nu_\mu$ and $\bar{\nu}_\mu$ scattering and include
the muon mass in our evaluation.
The label of {\it TMC} indicates that target mass corrections
can be incorporated into the structure functions \cite{gp,barbieri}, both
through the Nachtmann variable 
\begin{eqnarray}
\xi &=& \frac{2 x}{Q^2(1+\rho)}\\
\rho & = & (1+4 M_N^2x^2/Q^2)^{1/2}\ ,
\end{eqnarray}
with nucleon mass dependent factors multiplying the
asymptotic structure functions and with convolutions of
the asymptotic structure functions. 

At leading order in QCD, neglecting
target mass corrections and for values of $Q$ large enough that the
parton model makes sense, the structure functions are schematically
\begin{eqnarray}
F_2(x,Q^2)&=& 2x F_1(x,Q^2)\\ \nonumber
&=&\sum 2x \bigl(q(x,Q^2)+ \bar{q}(x,Q^2)\bigr)\\ \nonumber
F_3(x,Q^2) &=& \sum 2\bigl(q(x,Q^2) -\bar{q}(x,Q^2)\bigr)\ .
\end{eqnarray}
Only the relevant quarks and antiquarks are included above, and 
Cabibbo-Kobayashi-Maskawa mixing angles must be included as well.
The full expressions, including NLO corrections, target mass corrections
and charm mass corrections appear, e.g., in  Ref. \cite{dis}.

\begin{figure}[t]
\begin{center}
\epsfig{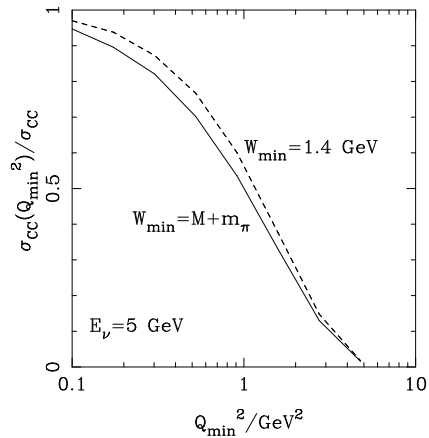}
\end{center}
\caption{$E_\nu=5$ GeV including NLO QCD corrections
and target mass corrections.
}
\label{fig:nlotmcrat5}
\end{figure}

QCD corrections reduce the DIS cross section by about 10\% for
$E_\nu=5-10$ GeV \cite{dis,kr}. The evaluation of the cross section
at these energies relies on an extrapolation of the parton distribution
functions to low $Q$. The parton
distribution functions are typically not defined below a minimum
factorization scale $Q_0^2$.
We set the PDF scale at $Q^2=Q_0^2$ for $Q^2<Q_0^2$ and
include the $O(\alpha_s)$ correction accounting for the mismatch of
factorization scale with the scale $Q$. Fig. \ref{fig:nlotmcrat5}
shows the ratio of the charged current neutrino-nucleon cross section
as a function of minimum $Q^2$ normalized to the total cross section for
$E_\nu=5$ GeV for two similar choices of $W^2_{\rm min}\sim 2\GeV2$. Approximately
half the cross section at this energy comes from $Q^2<1\GeV2$. At higher
energies, the fraction reduces, e.g. to $\sim 30\%$ at $E_\nu=10$ GeV.

The importance of $Q^2<1\GeV2$ in 
Fig. \ref{fig:nlotmcrat5} leads us to consider structure functions
at fixed values of $Q$ for a range of $x$ relevant to low energy
neutrino scattering. We focus on $F_2$ at $Q^2=0.1,\ 0.5, 1$ and $4 \GeV2$.
At $Q^2=4\GeV2$, we are firmly in the perturbative QCD regime, and at the
low end of $Q^2$ we are definitely out of the perturbative regime.
The range of $x$ at fixed
$Q^2$ is limited on the upper end by $W_{\rm min}^2$ as shown
in Fig. \ref{fig:wcontours}. The lower limit on $x$
is determined by the incident
neutrino energy:
$x\geq {Q^2}/({2ME_\nu})$. Given the plethora of data in electromagnetic
scattering, we turn to $ep$ scattering to test the perturbative evaluation
of the electromagnetic structure function $F_2$ and to consider alternatives. 

\section{$ep$ scattering}

Electron-proton scattering in the perturbative regime is well described
by the parton model. The structure function $F_2$
at leading order, neglecting target mass corrections,
is written in terms of quark (and antiquark) distribution functions
$q(x,Q^2)$ with electric charge $q=e_i e$:
\begin{equation}
F_2(x,Q^2)=\sum_ie_i^2 \Bigl( q(x,Q^2)+\bar{q}(x,Q^2)\Bigr)\ .
\end{equation}
As discussed above, the parton
distribution functions are extrapolated below the minimum
factorization scale $Q_0^2$ in the perturbative parton
model evaluation.

\begin{figure}[t]
\begin{center}
\epsfig{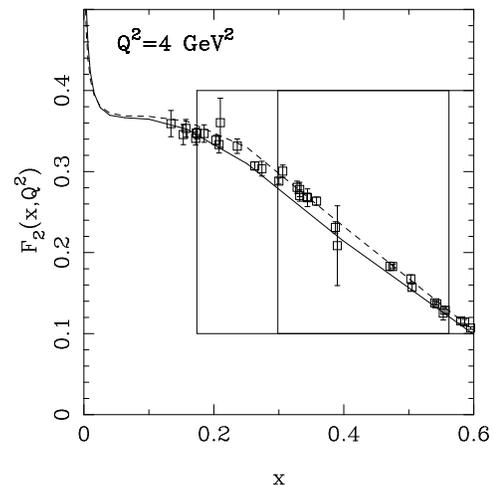}
\end{center}
\caption{$F_2(x,Q^2)$ for $ep$ electromagnetic scattering for $Q^2=4$ GeV, 
with the ALLM parameterization (solid), NLO QCD (dashed) evaluated
using the MRST2004. Also shown are representative data from Whitlow {\it et al.}
[25] for $Q^2=3.7-4.3\ \GeV2$.
}
\label{fig:f2q4}
\end{figure}

Structure functions calculated
in this manner can be compared with $ep$ electromagnetic
scattering data. For ease of comparison, we primarily use the
parameterization of Abramowicz, Levin, Levy and Maor \cite{allm}
which uses 23 parameters to describe a wide range of data. 
In Fig. \ref{fig:f2q4} we show the ALLM parameterization (solid line), along with
NLO QCD (dashed) evaluated using the Martin {\it et al.} 
parton distribution functions \cite{mrst}
MRST2004. The data points come from
SLAC $ep$ scattering data \cite{whitlow} for $Q^2=3.7-4.3\ \GeV2$.

Also in shown in Fig. \ref{fig:f2q4}
are boxes indicating the relevant range of $x$ for
$Q^2=4\GeV2$. The vertical line furthest left is $x_{\rm min}$ for
$E_\nu=10$ GeV, while the next vertical line is $x_{\rm min}$ for
$E_\nu=5$ GeV. The third vertical line shows $x_{\rm max}$ for
$W_{\rm min}^2=4\GeV2$. At larger $x$ is the maximum for $W_{\rm min}^2=2\GeV2$.
The perturbative calculation matches the ALLM parameterization and the
data well.

\begin{figure}[t]
\begin{center}
\epsfig{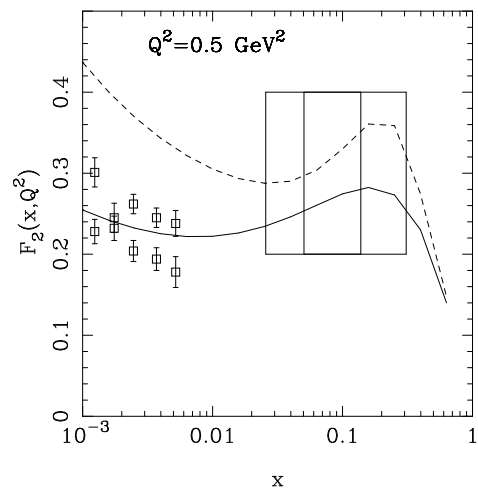}
\end{center}
\caption{$F_2(x,Q^2)$ for $ep$ electromagnetic scattering for $Q^2=0.5$ GeV, 
with the ALLM parameterization (solid), NLO QCD (dashed). The data are from the E665 experiment [26] with $Q^2=0.43\ \GeV2$
and $0.59 \ \GeV2$.}
\label{fig:f2qp5}
\end{figure}

For comparison, we show the same curves for $Q^2=0.5 \ \GeV2$ along
with data from E665 \cite{e665} in Fig.
\ref{fig:f2qp5}. The data are for
$Q^2=0.43,\ 0.59\GeV2$.  The pertubative evaluation of $F_2$ overestimates
the ALLM parameterization. 
On the basis of the discrepancy between $ep$
data and the perturbative curves, we conclude that the low energy
$\nu N$ DIS cross section is overestimated by the perturbative expression.

Two other phenomenological
parameterizations of electromagnetic structure functions
are discussed here. The first is by Capella, Kaidalov, Merino
and Thanh Van (CKMT) \cite{ckmt} and the second is
the Bodek-Yang-Park
parameterization \cite{by,byp}. 

\subsection{CKMT Parameterization}

The CKMT parameterization \cite{ckmt} is based on 
the form
\begin{eqnarray}
\nonumber
F_2(x,Q^2)&=& 
F_2^{sea}(x,Q^2)+F_2^{val}(x,Q^2) \\ \nonumber
&=&Ax^{-\Delta(Q^2)}(1-x)^{n(Q^2)+4}\Biggl(
\frac{Q^2}{Q^2+a}\Biggr)^{1+\Delta(Q^2)}\\ \nonumber
&+& Bx^{1-\alpha_R}(1-x)^{n(Q^2)}\Biggl( \frac{Q^2}{Q^2+b}\Biggr)^{\alpha_R}\\ 
&\times & \Bigl( 1+ f(1-x)\Bigr) \ .
\label{eq:ckmt}
\end{eqnarray}
The first term dominates at small $x$, where the physical picture
is that photons fluctuate into $q\bar{q}$ pairs that form
vector mesons. The second term dominates large $x$ and is interpreted
as the valence term. Characteristically, the valence $d$ quark
distribution has one extra power of $(1-x)$.

An interesting feature of this parameterization is its economy relative
to the ALLM parameterization. The quantities $B$ and $f$ are 
calculated by requiring two valence $u$ quarks and one valence
$d$ quark. This assumes that the first term
(proportional to $A$) has no valence content. $B$ is the coefficient
of the up valence component, and $f$ is the ratio of the down
valence to up valence coefficients.

The quantities $n(Q^2)$
and $\Delta (Q^2)$ are parameterized by CKMT according to the form
\begin{eqnarray}
n(Q^2)&=& \frac{3}{2}\Biggl( 1+\frac{Q^2}{Q^2+c}\Biggr)\ ,
\\
\Delta(Q^2)&=& \Delta_0 \Biggl( 1+\frac{2 Q^2}{Q^2+d}\Biggr)\ .
\end{eqnarray}
Values of the parameters from Ref. \cite{ckmt} appear in Table 1.

The quantity $\Delta_0\simeq 0.08$ represents the power
law of $F_2$ governed by pomeron exchange at low $Q^2$ \cite{dl}.
This is the same power law that appears in the generalized
vector meson dominance 
(GVDM) approach \cite{sakurai,bk}.
The $Q^2$ dependent prefactor, however, has the same form as the continuum
contribution rather than the vector meson contribution.
Alwall and Ingelman \cite{ingelman} have taken the GVDM approach
together with valence parton density functions based on a model
accounting involving quantum fluctuations to consider $ep$
scattering at low-$Q^2$. We take the more phenomenological CKMT
approach
to parameterizing $F_2$ in part because it is relatively straightforward
to convert to $\nu N$ structure functions.

\begin{table}
\caption{Parameter values in Ref. [13] for CKMT parameterization of
the electromagnetic structure function $F_2$. The quantities $B$ and $f$ are determined
from the valence conditions at $Q^2=2\ \GeV2$ rather than fit.} 
\begin{tabular}{lccc}
\hline
Parameter & Value & Parameter & Value [$\GeV2$]  \\
$A $ & 0.1502 & $a$ & 0.2631 \\
$\Delta_0$      &  0.07684 & $d$ & 1.1170 \\
$B$     &  1.2064 & $b$ & 0.6452 \\
$\alpha_R$       &  0.4250 & $c$ & 3.5489 
\\
$f$ & 0.15 & & \\
\hline
\end{tabular}
\end{table} 

\begin{figure}[t]
\begin{center}
\epsfig{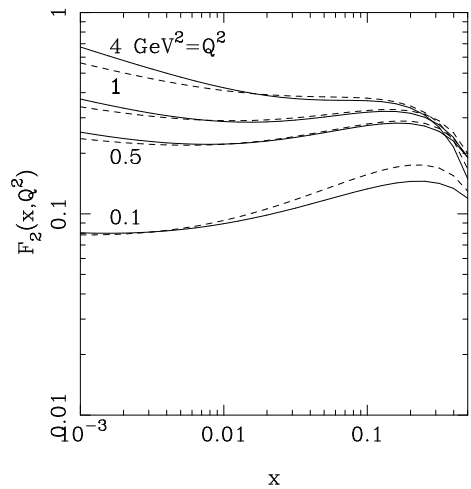}
\end{center}
\caption{$F_2(x,Q^2)$ for electromagnetic scattering  
with the ALLM parameterization (solid), and the CKMT parameterization
(dashed).}
\label{fig:ckmt}
\end{figure}

A comparison of the ALLM parameterization and the CKMT parameterization
of $F_2$ for electromagnetic scattering at $Q^2=0.1,\ 0.5, 1$ and $4\ \GeV2$
is shown in Fig. \ref{fig:ckmt}.

\subsection{Bodek-Yang-Park Parameterization}

A second parameterization has been performed by Bodek and Yang \cite{by},
joined by Park \cite{byp}. A generalized Nachtman variable $\xi_w$
is used,
along with form factors and the Gluck, Reya and Vogt parton distribution 
functions (GRV98) \cite{grv}so that the structure function $F_2$
is written as 
\begin{equation}
F_2(x,Q^2)\equiv\sum e_q^2 \xi_w (\tilde{q}(\xi_w ,Q^2)
\tilde{\bar{q}}(\xi_w ,Q^2)\ .
\end{equation}
The PDFs $\tilde{q}$ are related to the usual PDFs by
form factor rescaling. Details are given in the appendix.

This approach is specifically designed to be
used with neutrino scattering.  
One feature of the Bodek-Yang-Park (BYP)
analysis is the separation of the 
quark and antiquark flavors. The flavor separated components of $F_2$ 
are then used to evaluate
neutrino charged current interactions. To address the question of how
universal the BYP parameterizations of effective quark
distributions are, we will compare the BYP approach
to evaluating the $\nu N$ cross section with the CKMT
parameterization. A comparison of these two cross section results will
give an estimate of the uncertainty in the $\nu N$ DIS cross section
component. Ultimately, measurements in this kinematic regime
are required.

\section{Structure functions in neutrino scattering}

The evaluation of the neutrino structure functions using
the BYP parameterization are straightforward using 
\begin{equation}
F_2(x,Q^2)\equiv\sum 2 \xi_w (\tilde{q}(\xi_w ,Q^2)
\tilde{\bar{q}}(\xi_w ,Q^2)\ .
\end{equation}
The CKMT parameterization was fit specifically to $ep$
scattering data, however, with the interpretation of the
separate sea and valence terms, it can be modified to apply to
neutrino scattering. Here we consider only isoscalar
nucleons $N$, the average of proton plus neutron targets.

The modification of Eq. (\ref{eq:ckmt}) will be done only for
three parameters: $A,\ B$ and $f$ will be replaced by
$A_\nu,\ B_\nu$ and $f_\nu$ \cite{ckmtnu}. The rationale is that the small
$x$ and large $x$ structure
functions in electromagnetic and charged current
scattering should show the same qualitative behavior 
as a function of $Q^2$. At large $x$, 
the balance between valence up and down
quarks is different in electromagnetic and charged current
scattering, however, the coefficients $B$ and $f$ are calculable.
Qualitatively, for small-$x$, the structure functions are
sea dominated. Again, the mix of sea components is changed, but
the assumption is that each sea component has qualitatively
the same behavior in $x$ and $Q^2$, so only $A$ changes.

One caveat to this approach
is that at $Q^2=0$, PCAC requires
a modification of the functional form of the parameterization \cite{pcac}.
Neither the BYP nor the CKMT parameterization currently accommodates
this theoretical behavior. Nevertheless, the small $x$, small $Q^2$ data from neutrino scattering
are accommodated. For example, the CCFR data \cite{bonniethesis} 
for $x=0.008, 0.0125,0.0175$
for $Q^2=0.4-0.9\GeV2$ are within $\sim 10\%$ of the CKMT parameterization
with neutrino modifications discussed below.

The modifications to Eq. (\ref{eq:ckmt})
are the following. We calculate that
for neutrino scattering $B\rightarrow B_\nu=2.695$ and
$f_\nu=0.595$. By evaluating
$F_2$ at $Q^2=10\GeV2$, we choose $A\rightarrow
A_\nu=0.60$ so that
it matches the NLO-TMC corrected $F_2$ reasonably well. This gives
$A_\nu/A=4$, which is what one would estimate by counting sea
quarks and antiquarks contributing to each process, weighting
$s$ quarks and antiquarks by a factor of $1/2$. 

\begin{figure}[t]
\begin{center}
\epsfig{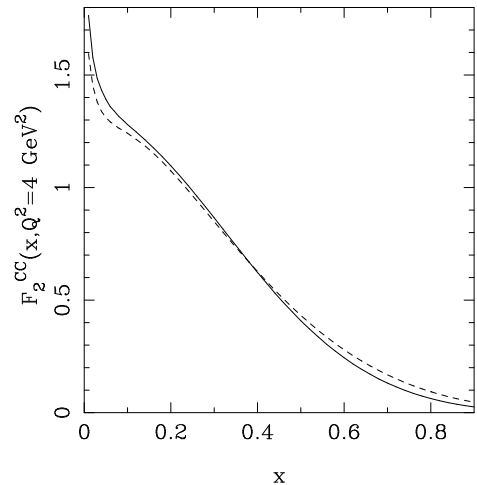}
\end{center}
\caption{Neutrino-isoscalar nucleon $F_2$ 
at $Q^2=4$ GeV$^2$ for charged current scattering,
solid line for NLO QCD with TMC, dashed line from Bodek-Yang-Park parameterization using GRV98.}
\label{fig:f2q24}
\end{figure}

Fig. \ref{fig:f2q24} shows a comparison of $F_2$ calculated
using GRV98 PDFS at NLO in QCD with target mass corrections
(solid line) and the Bodek-Yang-Park parameterization,
both at $Q^2=4\GeV2$. The two evaluations of $F_2$ agree well
at this value of $Q^2$.
At lower $Q^2$, the perturbative
evaluation is larger than the BYP parameterization, for example,
by about 30\% at $x=0.1$ and $Q^2=1\GeV2$.

\begin{figure}[t]
\begin{center}
\epsfig{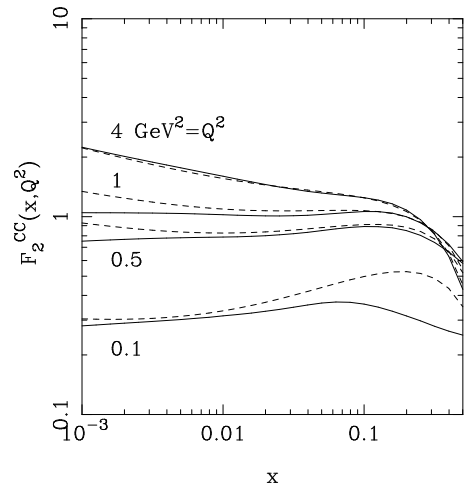}
\end{center}
\caption{Neutrino-isoscalar nucleon $F_2$ 
at $Q^2=0.1,\ 0.5,\ 1$ and 4 GeV$^2$ for charged current scattering,
solid line for the Bodek-Yang-Park parameterization using GRV98
and the dashed line for the modified CKMT parameterization 
with $A_\nu=0.60,\ B_\nu=2.695$ and $f=0.595$.}
\label{fig:f2ccckmt}
\end{figure}

Fig. \ref{fig:f2ccckmt} shows 
the charged current $F_2$ using the modified CKMT  (dashed lines)
and the BYP (solid lines)
parameterizations for several values of $Q^2$.
The parameterizations give similar results except
for the lowest value of $Q^2$ in the large $x$ region, and for the
smallest $x$ values.

Cross sections require $F_1$ and $F_3$ as well as $F_2$. 
For the CKMT parameterization of $F_3$, we start with the
valence term and add a strange quark component
equal to 1/15 of the total sea contribution in $F_2$.
To achieve a better large $x$ agreement with perturbative
QCD, the structure function has an overall
normalization factor of 0.91. This factor makes the
valence component integrate to 2.73 at $Q^2=2\GeV2$,
consistent with measurements \cite{pdg} of the Gross-Llewellyn-Smith
sum rule\cite{gls} including QCD corrections.
To summarize, we take
for neutrino scattering
\begin{eqnarray}
\nonumber
F_3(x,Q^2)&=&\Biggl[ \frac{A_\nu}{15} x^{-\Delta(Q^2)}(1-x)^{n(Q^2)+4}\Biggl(
\frac{Q^2}{Q^2+a}\Biggr)^{1+\Delta(Q^2)}\\ \nonumber
&+& {B_\nu} x^{1-\alpha_R}(1-x)^{n(Q^2)}\Biggl( \frac{Q^2}{Q^2+b}\Biggr)^{\alpha_R}\\ 
&\times & \Bigl( 1+ f_\nu (1-x)\Bigr)\Biggr]/(1.1 x) \ .
\label{eq:ckmt3}
\end{eqnarray}

$F_1$ is obtained from $F_2$ by including the
correction for $R\neq 0$ where
\begin{equation}
R=\frac{F_2}{2xF_1}\Biggl( 1+\frac{4M^2x^2}{Q^2}\Biggr) -1 \ .
\end{equation}
We 
use the parameterization of Whitlow et al. in Ref. \cite{whitlow},
which is also consistent with neutrino scattering data \cite{chorus}.
The parameterization applies for $Q^2>Q_m^2=0.3\GeV2$.
Below $Q^2=Q_m^2$, we take $R(x,Q^2)=R(x,Q_m^2)\cdot Q^2/Q_m^2$.

We note that similar results for the structure functions are obtained
by rescaling the NLO+TMC in a manner similar to Eq. (\ref{eq:ckmt}),
e.g., below the scale $Q_c^2$,
\begin{equation}
\bar{u}(\xi,Q^2)= \bar{u}(\xi,Q_c^2)\cdot F^{sea}_2(x,Q^2)/
F^{sea}_2(x,Q_c^2)
\end{equation}
and similarly for the valence distributions. The gluon distribution
must be rescaled according to the ``sea'' factor.

\section{Neutrino-Nucleon Cross Section}

The neutrino cross section with isoscalar nucleon targets is calculated
using Eq. (2). For the results labeled by the low-$Q^2$ parameterization,
we use the full NLO QCD corrected structure functions including target mass corrections for $Q^2>Q_c^2=4\GeV2$. Below
$Q_c^2$, we use either the BYP or CKMT parameterizations of
the nonperturbative region. 
For the
energies considered here, below $10$ GeV, the results are not very sensitive
to the cutoff $Q_c$. At the lowest energies, the nonperturbative
parameterization is relatively more important, however, the cross
sections are small.

Results labeled with NLO+TMC, as in previous sections, used
the next-to-leading order QCD corrected structure functions, with
parton distribution functions frozen at the minimum value provided
by the parameterization. For this section, all NLO+TMC results
used the GRV98 PDFs with a minimum $Q^2=0.8\GeV2$.

\begin{figure}[t]
\begin{center}
\epsfig{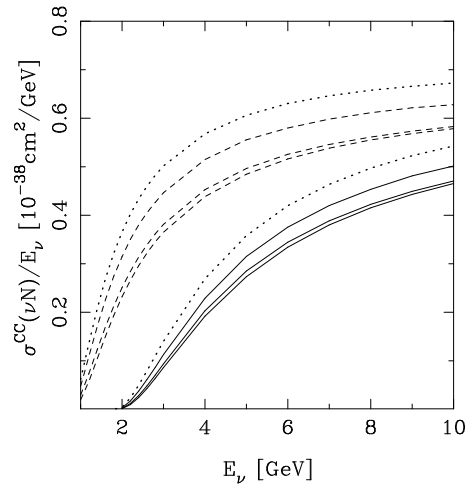}
\end{center}
\caption{Neutrino-isoscalar nucleon cross sections
normalized to incident neutrino energy: solid lines
for $W_{\rm min}^2=4\GeV2$ and dashed lines for $W_{\rm min}^2=2\GeV2$.
The upper solid and dashed lines use NLO QCD plus target mass corrections
to evaluate the cross section, while the lower solid and dashed lines
use the
CKMT  and BYP
parameterizations below $Q^2=4 \GeV2$. The dotted lines show the
evaluation using leading order QCD plus target mass corrections.}
\label{fig:ccrange}
\end{figure}

\begin{figure}[t]
\begin{center}
\epsfig{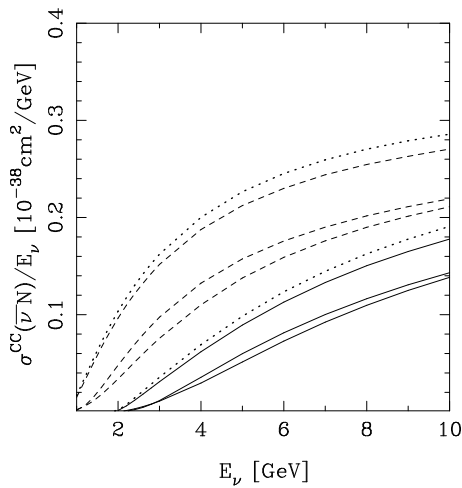}
\end{center}
\caption{Anti-neutrino-isoscalar nucleon cross sections
normalized to incident anti-neutrino energy: solid lines
for $W_{\rm min}^2=4\GeV2$ and dashed lines for $W_{\rm min}^2=2\GeV2$.
The upper solid and dashed lines use NLO QCD plus target mass corrections
to evaluate the cross section, while the lower solid and dashed lines
use the
CKMT  and BYP
parameterizations below $Q^2=4 \GeV2$. The dotted lines show the
evaluation using leading order QCD plus target mass corrections.}
\label{fig:accrange}
\end{figure}

In Fig. \ref{fig:ccrange} we show 
the neutrino-nucleon charged current cross section 
normalized by incident neutrino energy, where
the cross section have been evaluated using
NLO+TMC (upper lines)
and using the CKMT  and BYP parameterizations 
below $Q_c^2=4\GeV2$ (lower lines)
for $W_{\rm min}^2=4\GeV2$ (solid) and $2\GeV2$ (dashed).
The dotted lines show the cross sections calculated using
leading order QCD including target mass correction.
Numerical values corresponding to this figure are shown
in Table II. The corresponding results for antineutrino-nucleon
scattering are shown in Fig. \ref{fig:accrange} and Table III.

The low $Q^2$ corrections reduce the cross sections even at
$E_\nu=10$ GeV. For $W_{\rm min}^2=2\GeV2$ for neutrinos, the reduction
is by 7-8\% at 10 GeV, and 11-13\% at 5 GeV. For neutrino scattering
with $W_{\rm min}^2=4\GeV2$ the reduction ranges from 6-7\% to 10-15\%
for the same energies. Antineutrino scattering cross sections are
more dramatically affected. The cross sections are reduced by of
order 20\% for $E_{\bar{\nu}}=10$ GeV, and between 25-43\% depending
on $W_{\rm min}^2$ for 5 GeV incident antineutrinos. The sensitivity
to the value of $Q_c^2$ is greater, a few percent, at $E_{\bar{\nu}}=10$
GeV. 

\begin{table*}
	\centering
		\begin{tabular}{|l|ccc|ccc|}\hline
		$E_\nu$ [GeV] & NLO+TMC & BYP & CKMT & NLO+TMC & BYP & CKMT\\ 
		 & $W_{\rm min}^2=2\GeV2$& & & $W_{\rm min}^2=4\GeV2$& &\\ \hline
		 1 & 4.77e-2 & 1.96e-2 & 2.93e-2 & & & \\
		 2 & 6.29e-1 & 4.64e-1 & 5.02e-1 & 6.40e-3 & 2.83e-3 & 4.08e-3\\
		 3 & 1.34 & 1.10 & 1.14 & 3.38e-1 & 2.58e-1 & 2.81e-1\\
		 5 & 2.78 & 2.42 & 2.48 & 1.56 & 1.37 & 1.43 \\
		 10 & 6.28 & 5.78 & 5.83 & 5.02 & 4.65 & 4.71\\ \hline
		\end{tabular}
		\caption{Neutrino-nucleon charged current cross section, in 
		units of $10^{-38}$ cm$^2$, calculated using NLO QCD with target
		mass corrections, with the Bodek-Yang-Park parameterization and
		with the Capella et al. parameterization below $Q_0^2=4\GeV2$.}
\end{table*}
\begin{table*}
	\centering
		\begin{tabular}{|l|ccc|ccc|}\hline
		$E_{\bar{\nu}}$ [GeV] & NLO+TMC & BYP & CKMT & NLO+TMC & BYP & CKMT\\ 
		 & $W_{\rm min}^2=2\GeV2$& & & $W_{\rm min}^2=4\GeV2$& &\\ \hline
		 1 & 1.49e-2 & 2.17e-3 & 1.71e-3 & & & \\
		 2 & 1.93e-1 & 6.79e-2 & 9.58e-2 & 2.64e-3 & 4.43e-4 & 2.22e-4\\
		 3 & 4.55e-1 & 2.27e-1 & 2.92e-1 & 9.29e-2 & 3.29e-2 & 3.45e-2\\
		 5 & 1.06 & 6.89e-1 & 7.87e-1 & 4.46e-1 & 2.56e-1 & 2.99e-1 \\
		 10 & 2.70 & 2.11 & 2.19 & 1.78 & 1.38 & 1.43\\ \hline
		\end{tabular}
		\caption{Antineutrino-nucleon charged current cross section, in 
		units of $10^{-38}$ cm$^2$, calculated using NLO QCD with target
		mass corrections, with the Bodek-Yang-Park parameterization and
		with the Capella et al. parameterization below $Q_0^2=4\GeV2$.}
\end{table*}

The agreement between the cross sections calculated with the CKMT
and BYP parameterizations gives some confidence in the predictions
for the DIS component of the neutrino-nucleon cross section at
a few GeV in energy. This is one step in developing the comprehensive
calculation of the neutrino cross section required for 
interpreting the oscillation
measurements at present and in the future.

\vskip 0.2in
\noindent{\bf Appendix}
\vskip 0.2in

The form of the Bodek-Yang-Park parameterization \cite{by,byp} 
used in this
paper relies on parameterizations of the form
\begin{equation}
F_2(x,Q^2)\equiv\sum e_q^2 \xi_w \tilde{q}(\xi_w ,Q^2)
\end{equation}
where $\tilde{q}(\xi_w,Q^2)$ depends on rescaled Gluck, Reya
and Vogt PDFs \cite{grv}
in terms of a modified Nachtman variable $\xi_w$. For
massless final state quarks $\xi_w$ is used, while for
charm production, $\xi_{wc}$ is used instead. They are defined as
\begin{eqnarray*}
\xi_w &=& \frac{2 x(Q^2+B)}{Q^2(1+\rho)+2 A x}\\
\xi_{wc} &=& \frac{2 x(Q^2+B+m_c^2)}{Q^2(1+\rho)+2 A x}\\
A & = & 0.538\GeV2 \\
B &=& 0.305\GeV2\\
m_c & = & 1.5 \ {\rm GeV}\\
\rho & = & (1+4 M_N^2x^2/Q^2)^{1/2}
\end{eqnarray*}
The rescaling of the PDFs is of the form 
\begin{eqnarray*}
\tilde{u}_v&=&\frac{(1-G_D^2)\cdot(Q^2+C_{2vu})}{Q^2+C_{1vu}}u_v \\
\tilde{d}_v&=&\frac{(1-G_D^2)\cdot(Q^2+C_{2vd})}{Q^2+C_{1vd}}d_v \\
\tilde{\bar{u}}
&=& \frac{Q^2}{Q^2+C_{su}}\bar{u} \ ({\rm and\ sim.\ for\ }
\tilde{\bar{d}},\tilde{\bar{s}})
\end{eqnarray*}
where $G_D=(1+Q^2/(0.71\ {\rm GeV}^2))^{-2}$. The
valence and sea K factors are:
\begin{eqnarray*}
C_{1vu} &=& 0.291\GeV2\quad\quad
C_{1vd} = 0.202\GeV2\\
C_{2vu} &=& 0.189\GeV2\quad\quad
C_{2vd} = 0.255\GeV2\\
C_{su} &=& 0.363\GeV2\quad\quad
\ C_{sd} = 0.621\GeV2\\
C_{ss} &=& 0.380\GeV2\ .
\end{eqnarray*}
 
\noindent{\bf Acknowledgments}

This work was supported in part by DOE Contract DE-FG02-91ER40664.
The author thanks Y. Meurice, I. Sarcevic and
especially S. Kretzer for contributions and comments.

\end{document}